\documentclass[singlecolumn,superscriptaddress,10pt]{revtex4}

\usepackage{amsmath}
\usepackage{latexsym}
\usepackage{amsthm}

\usepackage{amssymb}
\usepackage{graphicx,epstopdf,color,hyperref,enumerate}
\usepackage{epsfig,amssymb,amsmath}

\def\>{\rangle}
\def\<{\langle}

\def\labell#1{\label{#1}}
\def\togli#1{}

\begin{document}

\title{\large Knowledge of Quantum Hidden Variables Enables Backwards-In-Time Signaling}

\author{Avishy Carmi}
\affiliation{Center for Quantum Information Science and Technology \& Faculty
of Engineering Sciences, Ben-Gurion University of the Negev, Beersheba
8410501, Israel}
\author{Eliahu Cohen}
\affiliation{Faculty of Engineering and the Institute of Nanotechnology and Advanced
Materials, Bar Ilan University, Ramat Gan 5290002, Israel}
\author{Lorenzo Maccone}
\affiliation{Dip.~Fisica and INFN Sez.\ Pavia, University of Pavia,
  via Bassi 6, I-27100 Pavia, Italy}
\author{Hrvoje Nikoli\'{c}}
\affiliation{Theoretical Physics Division, Rudjer Bo\v{s}kovi\'{c}
  Institute, P.O.B. 180, HR-10002 Zagreb, Croatia\\ $^*$Authors are listed
in alphabetical ordering.}


\begin{abstract}
  Bell's theorem implies that any completion of quantum
  mechanics which uses hidden variables (that is, preexisting values of
  all observables) must be nonlocal in the Einstein sense. This customarily
  indicates that knowledge of the hidden variables would permit
  superluminal communication. Such superluminal signaling, akin to the
  existence of a preferred reference frame, is to be expected.
  However, here we provide a protocol that allows an observer with
  knowledge of the hidden variables to communicate with her own causal
  past, without superluminal signaling. That is, such knowledge would
  contradict causality, irrespectively of the validity of
  relativity theory. Among the ways we propose for bypassing the
  paradox there is the possibility of hidden variables that change
  their values even when the state does not, and that means that
  signaling backwards in time is prohibited in Bohmian mechanics.
\end{abstract}
\maketitle

\section{Introduction}

In this paper we provide a protocol that allows any party with
knowledge of quantum hidden variables to communicate to her own past,
leading to a breakdown of causality in the strong sense.  By
``causality'' here we simply mean that ``a cause temporally precedes
all its effects'' (in the absence of closed timelike curves
\cite{goedel}). We will specify better the distinction between
``cause'' and ``effect''  in Sec.~\ref{s:causality}.  By ``hidden variables'' here
we mean ``whatever information that is necessary to describe values of
observables as preexisting before the measurement'', which is their
usual definition. Namely, the hidden variables encode the properties
that are assigned to the outcome once the measurement is performed.

Our protocol is based on quantum contextuality, and the context is
chosen {\em after} the values of the hidden variables are discovered.
But, since the context determines the value of the hidden variables
and different contexts will lead to different values, this value must
have been ``sent back in time''.  This applies to contextual hidden
variables. We will not consider non-contextual hidden variables, since
it is known that non-contextual hidden variables cannot be used to
describe quantum mechanics \cite{ks,peresmermin}.

In other words, the {\em future} choice of context affects the {\em
  past} values of the hidden variables, and this can be used to send
information back to one's own causal past as we shall show below.

Bell's theorem \cite{bell} implies that such hidden variables are
nonlocal: a measurement on one system may change the hidden variables
of a distant second system that may be causally disconnected from the
first. If the hidden variables exist, this type of nonlocality (which
we will call ``Einstein nonlocality'') is incompatible with the
conjunction of special relativity and causality. Indeed, a person with
knowledge of the hidden variables will be able to communicate
superluminally \cite{werner}, which means that {\em for another
  observer} the temporal order of cause and effect can be flipped. An
example can illustrate this: suppose that Alice uses a superluminal
projectile (or signal) to shoot a distant target.  The other observer,
Bob, in uniform motion with respect to Alice, can see these two
spacelike-separated events with opposite time ordering: he can see the
target is hit before Alice pulls the trigger. Interestingly, we note
that superluminal signals are not, by themselves, incompatible with
relativity if one relaxes causality \cite{sonego}. If Alice is a mafia
hit-man, she will be acquitted from killing her target thanks to Bob's
testimony that the target died before she even pulled the trigger. In
other words, the three hypotheses (i)~``special relativity'',
(ii)~``weak causality'' (a cause temporally precedes an effect) and
(iii)~``superluminal signaling'' lead to a contradiction: either one
has to drop causality (namely, Alice walks free), or one has to drop
superluminal signaling (these projectiles do not exist), or one has to
drop special relativity.  In other words, Bell's theorem implies that
the knowledge of hidden variables violates the causal relations of
another observer, under the assumption that special relativity holds.

In this paper we prove a stronger statement that one can communicate
with one's {\em own} past without any assumption about relativity.
Namely, we show that knowledge of contextual hidden variables enables BITS (Backwards-In-Time
Signaling). This is a stronger statement as is evident from the
observation that Newtonian mechanics is a theory that does allow
superluminal communication but does not allow any communication with
one's own past. Also a version of quantum mechanics where the collapse
of the wave function happens instantaneously in some preferred
reference frame will not allow one to communicate to one's own
past\footnote{ If one assumes relativity, superluminal communication
  by itself may allow one to communicate to one's own past, under the
  hypothesis that an observer can communicate to spacelike separated
  points that are in the ``future'' with respect to the observer's
  simultaneity hyperplane. Indeed, if Alice communicates
  superluminally a signal to Bob, Bob then accelerates to a large
  fraction of $c$ so that his hyperplane of simultaneity intersects
  Alice's timeline at a time before she sent the signal and then he
  resends Alice's signal along such hyperplane (superluminally in
  Bob's new reference frame), then Alice receives her signal back
  before she sent it.}.

Retrocausality, as well as weaker variants of the BITS property of ontic
quantum mechanical models have been pointed out in the literature
under additional hypotheses, e.g.  when considering fundamentally
time-symmetric versions of quantum mechanics (i.e.~even in the
presence of collapse) \cite{rod,price1,price2,price3,leifer,wharton,sandu,aharonov}.
However, when addressing Bell tests such retrocausal models only assume the existence of hidden variables that will be measured in the future. Hence these arguments assume communication to the past from the onset. Here,
instead, we use plain textbook quantum mechanics together with the
hypothesis that one could somehow know quantum hidden variables
(as defined above) to prove they must be BITS. Nevertheless, the aforementioned retrocausal frameworks showcase the existence of BITS models which are consistent with standard approaches to quantum mechanics since they strictly prohibit knowledge of the hidden variables.

An important interpretation of quantum mechanics that uses hidden
variables is Bohmian Mechanics. In Bohmian Mechanics spin hidden
variables do not exist, and the only hidden variables are the
positions of particles \cite{zanghi}, however we show below in which
sense our argument also applies to this case.

\section{The protocol}
Consider the Peres-Mermin square \cite{peresmermin,mermin} composed of the operators
\begin{align}
\begin{bmatrix}
\sigma_{1x}&\sigma_{2x}&\sigma_{1x}\sigma_{2x}\cr
\sigma_{2y}&\sigma_{1y}&\sigma_{1y}\sigma_{2y}\cr
\sigma_{1x}\sigma_{2y}&\sigma_{1y}\sigma_{2x}&\sigma_{1z}\sigma_{2z}
\end{bmatrix}
\labell{pmsq}\;
\end{align}
where $\sigma_{1j}$ are the operators that refer to the spin
components of a first spin-1/2 particle and $\sigma_{2j}$ of a second
spin-1/2 particle, both located in Alice's lab.  In Peres' words:
``The three operators in each row and each column commute, and their
product is $+1$, except those of the last column, whose product is
$-1$.  There is obviously no way of assigning numerical values $\pm 1$
to these nine operators with this multiplicative property''
\cite{peresmermin}. This implies that we cannot assign definite values
to the spin components of the two particles that are independent of
what is (was, will be) measured on the other particle: a very simple
consequence of quantum contextuality of hidden variables. Since
non-contextual hidden variables are impossible
\cite{peresmermin,mermin,ks}, one must conclude that the values of the
hidden variables of the spins must be fixed by the context, which is our assumption here.  The idea
of our protocol is to delay the choice of the context to a time
successive to when the values of the hidden variables have been used to
determine some ``signal''. This can be achieved through the protocol
composed by the following steps (see Fig.~\ref{f:fig}):

\begin{figure*}[hbt]
\centering

\includegraphics[width=0.95\textwidth]{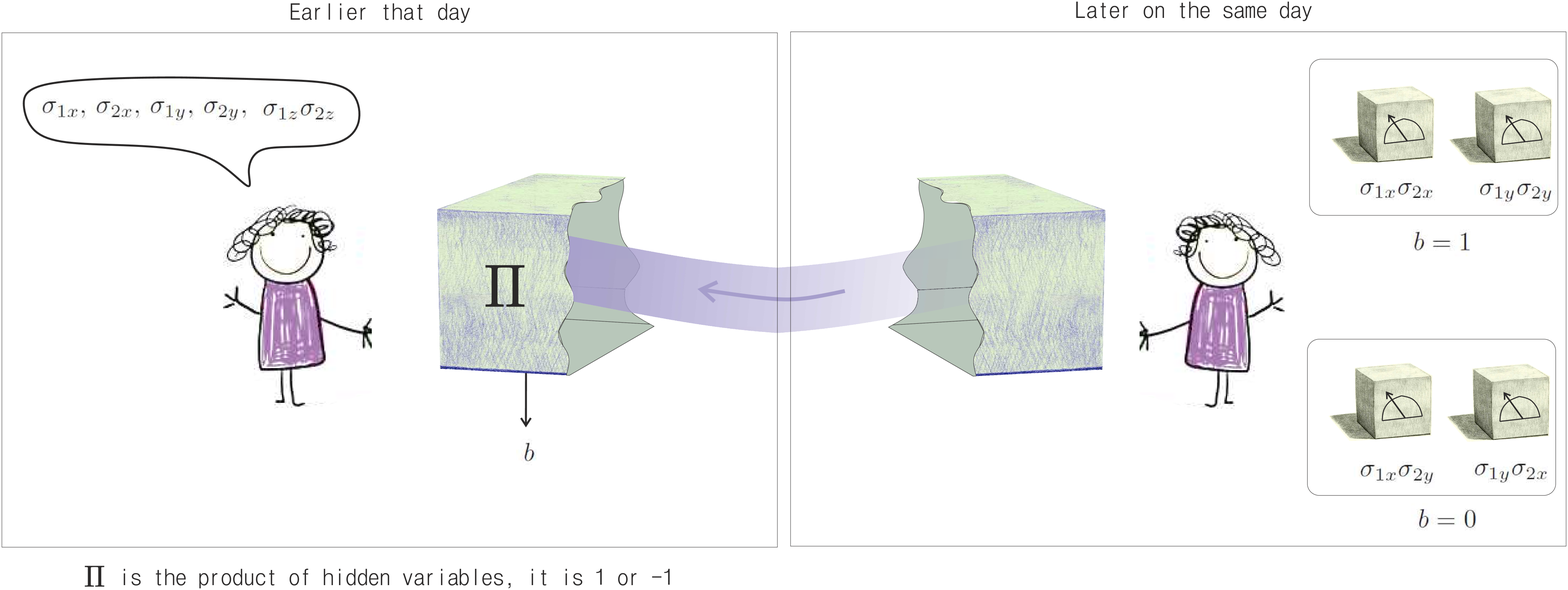}

\caption{Depiction of the protocol. First Alice finds out the 
  values of the (contextual) hidden variables and uses them to calculate
  the value of a bit $b$. Later she establishes the context by
  choosing which set of observables to measure. Her later choice
  determines the value of $b$, sending it back in time.}
\labell{f:fig}
\end{figure*}

\begin{enumerate}
\item Alice finds out the (contextual) hidden values of $\sigma_{1x}$,
  $\sigma_{2x}$, $\sigma_{1y},$ $\sigma_{2y}$,
  $\sigma_{1z}\sigma_{2z}$. Of course quantum mechanics prevents us
  from knowing these values because they refer to non-commuting
  (complementary) properties. However, the explicit assumption here is
  that Alice can know the hidden variables. As proven by Peres, the
  hidden variables are contextual, but the context will be established
  by Alice in the future.  Clearly we suppose that the values of the
  hidden variables reflect the values that are (or were or will be)
  uncovered by the experiments that are (or were or will be)
  performed. As specified above, this is the minimal requirement for
  ``hidden variables''. We emphasize the distinction between attaining
  knowledge of hidden variables and performing a quantum measurement.
  If one knew the hidden variable, one could predict the measurement
  outcome. Here we assume that one could somehow know the hidden
  variables {\em without} performing a quantum measurement. All
  interpretations of quantum mechanics prevent the knowledge of hidden
  variables (hence the name), but here we explore what would happen if
  it were possible to know them: if one assumes they exist, it is
  natural to ask what would happen if they could be known.
\item\label{s2} Consider the products in the table \eqref{pmsq}: take
  the product of the hidden values of $\sigma_{1x}$, $\sigma_{2x}$,
  $\sigma_{1y},$ $\sigma_{2y}$, and $\sigma_{1z}\sigma_{2z}$. If this
  product is equal to $+1$, then set a bit $b=0$, if the product is
  equal to $-1$, then set $b=1$. The first choice is consistent with
  looking at the last line of the Peres-Mermin square, the second
  choice with looking at the last column.
\item\label{s3} Then in the (distant) future, Alice can retroactively
  decide the value of $b$ by deciding what to measure. If she measures
  $\sigma_{1x}\sigma_{2x}$ and $\sigma_{1y}\sigma_{2y}$ (which
  commute), then she sets $b=1$. Since the hidden variables, by
  assumption, reflect the preexisting values of the observables, this
  means that by setting $b=1$ she sends back to her past self $b=1$.
  Instead, if she measures $\sigma_{1x}\sigma_{2y}$ and
  $\sigma_{1y}\sigma_{2x}$ (which also commute), then she sends back
  to herself $b=0$. Note that quantum complementarity forces Alice to
  choose one of these two possibilities: She cannot perform the
  measurements connected to both choices since they do not commute.
\end{enumerate}

The measurements in point \ref{s3} of the protocol can be performed in
the following way. The joint measurement of $\sigma_{1x}\sigma_{2x}$
and $\sigma_{1y}\sigma_{2y}$ can be performed through a measurement
that has Bell states as eigenstates. Indeed, consider $|a\>,|b\>$
eigenstates of $\sigma_x$ and $|\pm\>=(|a\>\pm|b\>)/\sqrt{2}$
eigenstates of $\sigma_y$, then
\begin{align}
|aa\>+|bb\>=|{+}{+}\>+|{-}{-}\>\!:  \sigma_{1x}\sigma_{2x}=+1,\;
\sigma_{1y}\sigma_{2y}=+1\nonumber&&\\
|aa\>-|bb\>=|{+}{-}\>+|{-}{+}\>\!:  \sigma_{1x}\sigma_{2x}=+1,\;
\sigma_{1y}\sigma_{2y}=-1\nonumber&&\\
|ab\>+|ba\>=|{+}{+}\>-|{-}{-}\>\!:  \sigma_{1x}\sigma_{2x}=-1,\;
\sigma_{1y}\sigma_{2y}=+1\nonumber&&\\
|ab\>-|ba\>=|{+}{-}\>-|{-}{+}\>\!:  \sigma_{1x}\sigma_{2x}=-1,\;
\sigma_{1y}\sigma_{2y}=-1\nonumber&&\\\label{ss1}
\end{align}
Analogously, the joint measurement of $\sigma_{1x}\sigma_{2y}$ and
$\sigma_{1y}\sigma_{2x}$ can be performed by the measurement that has
the following two-spin maximally entangled states as eigenstates:
\begin{align}
|a{+}\>+|b{-}\>=|{+}a\>+|{-}b\>\!:  \sigma_{1x}\sigma_{2y}=+1,\;
\sigma_{1y}\sigma_{2x}=+1\nonumber&&\\
|a{+}\>-|b{-}\>=|{+}b\>+|{-}a\>\!:  \sigma_{1x}\sigma_{2y}=+1,\;
\sigma_{1y}\sigma_{2x}=-1\nonumber&&\\
|a{-}\>+|b{+}\>=|{+}a\>-|{-}b\>\!:  \sigma_{1x}\sigma_{2y}=-1,\;
\sigma_{1y}\sigma_{2x}=+1\nonumber&&\\
|a{-}\>-|b{+}\>=|{+}b\>-|{-}a\>\!:  \sigma_{1x}\sigma_{2y}=-1,\;
\sigma_{1y}\sigma_{2x}=-1\nonumber&&\\\label{ss2}
\end{align}
Both of these observables are incompatible (do not
commute) with the measurement of the single spins, represented by the
observables $\sigma_{1x}\otimes\openone_2$,
$\openone_1\otimes\sigma_{2x}$, etc. Thus, textbook quantum mechanics
does not lead to paradoxes, namely either the local values are defined
{\em or} the joint values are determined by the measurements defined in
\eqref{ss1}-\eqref{ss2}. The Peres-Mermin argument shows that an
apparent paradox arises if one can determine both the local values
(through the spin hidden variables) and the joint values. Our protocol
hinges on this.

\section{Ways to bypass the argument?}
Here we consider two ways in which our argument can be apparently
bypassed: either by introducing additional hidden variables, or by
introducing a time dependence of the hidden variables.

The first possible objection is that one could argue that the hidden
variable for the joint measurements are unrelated to the hidden
variables of the local measurements. Namely that the hidden variable
of the product $\sigma_{1x}\sigma_{2x}$ is different from the product
of the hidden variables $\sigma_{1x}\cdot\sigma_{2x}$, but
fundamentally, not only statistically (at the level of expectation
values) \cite{vaidman} and not necessarily within a pre- and
post-selected ensemble \cite{vaidman1,vaidman2}.  \togli{The product
  hidden variables can be called ``pseudolocal'' \cite{pseudolocal} (a
  connection between them will lead back to local correlations as
  shown in the Appendix). }From the abstract point of view, this may
be an acceptable objection, but it leads to an implausible ontology.
Indeed, the observables connected to the product
$\sigma_{1x}\otimes\sigma_{2x}$ and the single values
$\sigma_{1x}\otimes\openone_2$ and $\openone_1\otimes\sigma_{2x}$ all
commute, namely they can be determined at the same time. Clearly when
this is done, the results {\em must} agree in any sensible ontology.
Indeed consider a scenario where two people may possess either a
yellow or a red apple, and suppose they both have a yellow apple
(namely the ``value'' of their apple is $+1$). Yet, if one assumes
that the value of the product is different from the product of the
values, they may conclude that their apples have different color
(namely the product of their ``values'' is $-1$), a nonsensical
conclusion. The fact that the measurements in \eqref{ss1} and
\eqref{ss2} are incompatible with the local measurements is irrelevant
to this argument, since it refers to the ontological ``true'' values,
not to the way these are measured (namely the context, which is
determined only in the future).

A second possible objection is that one may postulate that the
measurement itself is changing the hidden variables, namely that the
measurements of \eqref{ss1} and \eqref{ss2} are not constrained by the
hidden values of the spins. This indeed removes the possibility of
using Alice's choice of context to communicate back in time. However,
it also depletes the meaning of hidden variables. If the measurements
can change the hidden values, then the hidden values do not
pre-determine the measurement outcomes, namely the hidden variables do
not encode the ``true'' value of the property uncovered by the
measurement. Then they would be completely useless, and one could just
as well claim that there is {\em no} true value and the measurement
outcome is ``created'' at the time of the measurement, as in the
Copenhagen interpretation of quantum mechanics. In other words, if one
claims that the choice of the context may change the values of the
hidden variables {\em at the time of the choice}, one cannot describe
situations (such as the ones of our protocol) where the context is
deliberately left undetermined until  the values of the hidden
variables have been used. And this renders hidden variables
superfluous.

In conclusion, to bypass our argument either one needs to introduce a
nonsensical ontology where two identical apples are different, or one
must deplete the meaning of hidden variables.

\section{Bohmian mechanics}

In the de Broglie-Bohm interpretation of quantum mechanics \cite{dbb}
spins do not have hidden variables: the only hidden variables are the
particles' positions, but our argument employs spin hidden variables.
Nonetheless, we can easily translate the spin degrees of freedom into
trajectories.  Consider a ``full loop Stern-Gerlach apparatus'' (FULO)
where we do not determine whether the particle goes up or down, but
where we just subject the particle to a magnetic field gradient and
then we use the opposite magnetic field gradient to re-merge the two
arms of the wave function (e.g.~see \cite{folman}). Such an apparatus does not
measure anything: it is described by an identity transformation in
quantum mechanics, as it is a unitary transformation followed by its
inverse unitary. Nonetheless, the spin information is transferred into
the trajectory information inside the FULO. To determine the $x,y,z$
spin hidden variables, Alice can perform one FULO along the $x$ axis
followed by one FULO along the $y$ axis and one FULO along the $z$
axis and track (but not measure!)  the internal trajectories. \togli{In
principle, knowing without measuring is not absolutely impossible in
Bohmian mechanics. After all one can always simulate Bohmian mechanics
with specified initial conditions on a computer.}

These trajectories encode the spin ``hidden variables'' because the
trajectories in the FULOs are consistent when repeated: if an
$x$-directed FULO is followed by a $y$-directed one and then by a
second $x$-directed one, the trajectories along the two $x$-directed
ones coincide. Namely, any time dependence of the trajectories is
irrelevant in the encoding of the spin hidden values into FULOs
trajectories. Moreover, the FULOs trajectory encodes the outcome of
future spin measurements (as requested by the hidden variable
definition we use in this paper), since the measurement trivially
consists in measuring in which of the two arms the particle is
present, and since the trajectories in successive FULOs match.  Then,
we can conclude that the trajectory in the $w$-directed FULO encodes
the $w$ hidden value of the spin for $w=x,y,z$. Namely, the
``non-existent'' hidden variables for spin are translated into hidden
variables for position.

Then our argument {\em seemingly} applies: by knowing (not measuring!)
the trajectories of the particles through the three successive FULOs,
Alice seemingly would be able to signal back in time. Indeed, we have
shown how Alice can map the spin quantum degree of freedom into
position degrees of freedom of a particle which are the hidden
variables of Bohmian mechanics.

However, in reality our argument is not applicable to Bohmian
mechanics because of the time evolution of the hidden variables. Since
only position is a hidden variable in Bohmian mechanics, one must also
postulate that positions {\em at different times} may be connected to
different (incompatible) hidden variables. In our protocol we must
postulate that the successive FULOs refer to the spin hidden variables
{\em at the same time}. While this is a reasonable postulate, it is
false.  It is reasonable because the FULOs are just identity
transformations for quantum mechanics and they do not change the
quantum state at all, so one would expect that they cannot change the
hidden variables that encode the ``true'' value of the spin (as
confirmed by the fact that the trajectories match in successive FULOs
oriented along the same directions). However, it is false because if
one looks at the trajectory across two FULOs that are rotated by 180
degrees, one finds a trajectory inconsistent with the hidden variable
(see Fig.~\ref{f:fulo}). Namely, if one looks at the trajectory along
a FULO oriented along the $x$ direction and then at the trajectory
along a flipped FULO oriented along $-x$, one must conclude that the
hidden variable for $\sigma_x$ is flipped by the second FULO.

\begin{figure}[hbt]
\includegraphics[width=0.45\textwidth]{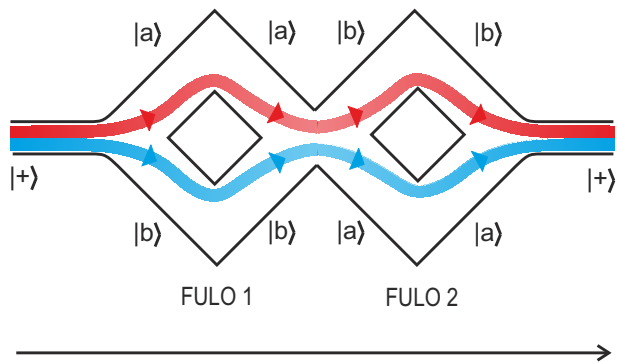}
  \caption{FULOs change the hidden variables.  A full-loop
    Stern-Gerlach apparatus (FULO) maps the spin hidden variables onto
    a particle trajectory. However, the FULO does not preserve the
    values of the hidden variables as shown by the conceptual
    experiment depicted here, where a FULO oriented along the $x$ axis
    is followed by a FULO oriented opposite to the $x$ axis. If the
    initial state of the spin is an equal superposition of ``up'' and
    ``down'' along the $x$ axis, namely $|+\>=(|a\>+|b\>)/\sqrt{2}$,
    then the trajectory is as depicted: if the trajectory in the first
    FULO is in the $a$-arm, it is in the $b$-arm in the second (which is flipped by
    180 degrees) and vice versa. A possible interpretation is that the
    FULO has changed the spin hidden variable for $x$, even though it
    is an identity transformation that does not change the quantum
    state.  }
\labell{f:fulo}
\end{figure}

In other words, Bohmian mechanics cannot be used for signaling back in
time by our protocol because Bohmian mechanics is an explicit
realization of the second way to bypass the argument that we discussed
in the previous section. The time dependence of the hidden variables
is nontrivial: even unitary evolutions may change the hidden variables
of spins. Therefore associating hidden variables with spins (and other
observables except for positions) is superfluous in Bohmian mechanics
\cite{NEW-REF}.

\section{\label{s:causality}Causal relations between past and future events}
In this section we elaborate more carefully on how one can distinguish
a ``cause'' event from an ``effect'' event. 

Indeed, a nontrivial subtlety is implicit in all protocols claiming to send
information back in time: such protocols establish a correlation
between two variables, in the past and in the future, and one must
prove that it is the future value that is ``causing'' the past one and
not vice versa (since obviously, correlation does not imply
causation). In other words, we have to establish without doubt that
the above protocol is sending the value of $b$ to the past, rather
than trivially asserting that the past value of $b$ determines the
future one. This can be established, but it comes at the price of
further assumptions. Different, {\em alternative}, assumptions are
possible and are enumerated here:

\begin{enumerate}
\renewcommand{\theenumi}{\Alph{enumi}}
\item Undetermined choice. In the above narrative we supposed that in
  the future Alice will be ``free'' to choose whatever value of $b$
  she wants, namely that her choice is not determined by the value of
  $b$ that was found in the past. This is tricky, since whatever her
  (free) choice, it will indeed be equal to the value of $b$ that was
  found in the past: that value is what she will choose to send (her
  choice is already known before she makes it). This is the very
  meaning of communication to the past. It implies that, even though
  she is free to choose whatever $b$ she wants, she cannot choose to
  send the value opposite to the one that was found in the past:
  whatever value is found in the past is the one she will (freely)
  choose in the future. While it is not strange that the value of her
  choice is known {\em after} she chose it, because of BITS, the value
  of her choice is known even {\em before} she chose it. 
    Causality here is somewhat confusing because the temporal order of
    events is reversed with respect to usual expectations. Therefore, asking
    what Alice will feel if she decides to send back to the past the
    value opposite of $b$ is meaningless, since she cannot decide to
    send back the opposite of what she actually decided to send. Sometimes
  what we called ``undetermined choice'' is termed ``free will'', but
  the adherents to compatibilism have long argued that free will is
  not inconsistent with determinism (e.g.~\cite{seth}), whereas here
  we simply require Alice's choice to be not pre-determined.  Because
  the ``free will'' and ``undetermined choice'' are quite slippery
  concepts especially for those interpretations that consider quantum
  mechanics a deterministic theory, one would possibly want to avoid
  physical consequences being attached to this hypothesis, so we will
  list other three that can {\em replace} it. Namely, the
    following three hypotheses are compatible with a ``no free will''
    condition according to which Alice is not ``free'' to choose the value of
    the bit, which is determined by some other process.
\item Evolutionary principle. This principle states that ``knowledge
  comes into existence only through evolutionary processes''
  \cite{deutsch}. This means that complex meaningful information (such
  as a Renaissance painting or a physics textbook) does not appear
  instantaneously from a random fluctuation, but it is the result of a
  lengthy evolutionary process or computation. For example, a
  Neolithic caveman could not have had the knowledge or the technique
  to paint Leonardo's Mona Lisa painting, and Newton could not have
  had the knowledge to write a quantum mechanics textbook. This
  implies that by tying the transmitted bit to the result of a long
  evolutionary process, one is guaranteed that the value of the bit
  cannot have been already known in the past. For example, we could
  tie the value of $b$ to some information that no one knows today,
  such as ``will Apple stock shares increase in value in ten years?''.
  This guarantees that indeed the bit was transmitted, and not known
  beforehand. Communication with the past together with the
  evolutionary principle lead to a chronology paradox \cite{deutsch}:
  suppose that Alice sends to Leonardo a picture of his painting and
  he painted it by copying her picture, then the painting would be
  generated spontaneously since Alice obtained it from Leonardo and
  Leonardo from Alice.
\item Relativistic causality (strictly enforcing non-superluminal
  communication of at least one event). A completely independent way
  of ensuring that $b$ is sent to the past is to assume relativity and
  causality (as defined above) and tie the value of $b$ to the unknown
  value of some degree of freedom that is spacelike separated from the
  observer at step \ref{s2} of the above protocol, but is accessible
  at step \ref{s3} of the protocol when one must ``choose'' the value
  of $b$.  For example: ``has the star Betelgeuse turned into a
  supernova?''  (Betelgeuse, in Orion, is a supernova candidate
  \cite{harv}). The supernova event is spacelike separated from step
  \ref{s2} and hence inaccessible if relativity plus causality is
  assumed, but it will become accessible at step \ref{s3}, when it
  enters in step \ref{s3}'s past lightcone. A caveat is in order here.
  As discussed above, relativity + causality is incompatible with
  superluminal communication, but the knowledge of hidden variables
  immediately implies superluminal communication. So one can assume
  relativity + causality for the event that causes the value of $b$
  (e.g.~the supernova explosion) {\em only} if one is sure that the
  event in question is not communicated superluminally through some
  quantum hidden variables.
\item Ontic randomness for some hidden variable. This entails that, at
  least in some cases, the outcomes of quantum measurements are
  intrinsically random. This is what the Copenhagen interpretation of
  quantum mechanics suggests, but an interpretation that uses hidden
  variables could say that randomness only arises because of lack of
  knowledge of the hidden variables. Indeed, hidden variables are
  based on the idea that a definite, pre-determined value existed
  prior to the measurement.  So, to be really intrinsically random,
  the hidden variables themselves must possess (at least in some
  cases) an ontic intrinsic randomness. Under this hypothesis, Alice
  may perform a $\sigma_z$ measurement of a qubit in an eigenstate of
  $\sigma_x$ using an ancillary qubit whose hidden value is not
  predetermined.  She then sends back the outcome as the value of $b$,
  which could not have been determined in the far past under this
  hypothesis. \end{enumerate}

Any of the above four hypotheses can be employed to conclude that in
the above protocol the communication proceeds from the future to the
past and not vice versa.

\section{Conclusions}

All interpretations of quantum mechanics that rely on hidden
variables, such as the de Broglie-Bohm theory \cite{dbb}, have some
kind of censorship mechanism which prevents the revelation of the
values of the hidden variables in practice, e.g.~see \cite{hvroje}.
Hence, one may claim that our protocol is of no practical relevance.
But even without such censorship, previous investigations on whether
Bohmian mechanics implies change of the past hidden variables in the
context of quantum erasure concluded that ``there is no change of the
past whatsoever'' \cite{fankauser}.  Nevertheless, the protocol
presented in our paper reaches the opposite conclusion for a wide
class of hidden variables by making use of quantum contextuality.
While we agree with previous findings that there is no change of the
past in Bohmian mechanics, our results show that it is misleading to
think of Bohmian mechanics as a ``hidden variables'' theory because
defining hidden variables for all observables except positions is in
Bohmian mechanics superfluous.  From an instrumental point of view,
Bohmian mechanics is best viewed as a theory of positions of
macroscopic objects explained in terms of microscopic positions of
their constituents, without any sharp borderline between
``non-hidden'' macroscopic positions and the ``hidden'' microscopic
ones \cite{hvroje}.

In previous works \cite{a,b,c} we have seen that quantum uncertainty is vital for reconciling quantum nonlocality in space and time with relativistic causality. And indeed, full access to the hidden variables in our protocol renders quantum uncertainty ineffective and thus it cannot circumvent the BITS.

One can then take the implications of our argument as further evidence
that the censorship mechanism that prevents the knowledge of the
hidden variables must be strongly enforced if one wants to preserve
chronology. In other words, knowing the hidden variables not only
allows for superluminal communication, but it even allows for
communication to the past. As explained above, this is a stronger
statement than simply requiring that Einstein locality (i.e.~no
superluminal communication) is satisfied. Of course, if hidden
variables are unknowable as a matter of principle, they can then be
considered unphysical for all intents and might be relegated to the
realm of metaphysics as ``unperformed experiments have no results''
\cite{peresbook,peresun}. And that is the core implication of this
paper.

In essence, our conclusion is that in contextual (fully deterministic)
hidden variables models for quantum theory, where agents have
underdetermined choice and can know the values of the hidden
variables, communication to one's own causal past would be possible,
negating the non-relativistic notion of causality. The same conclusion
holds also under slightly broader hypotheses that do not require
undetermined choices on the part of an agent, as discussed in
Sec.~\ref{s:causality}.\\[4ex]

\noindent
{\bf Data accessibility.} This paper has no data.\\[1ex]
{\bf Competing interests.} The authors declare no competing
interests.\\[1ex]
{\bf Authors' contributions.} L.M. conceived the main idea which was later developed by all
  authors. All authors gave final approval for publication.\\[1ex]
{\bf Acknowledgements.} We would like to thank Ken Wharton for helpful comments and insightful discussions. \\[1ex]
{\bf Funding statement.}
L.M. acknowledges funding from Unipv ``Blue sky''
project - grant n.~BSR1718573 and the FQXi foundation grant
FQXi-RFP-1513 ``the physics of what happens''. HN acknowledges funding
by the Ministry of Science of the Republic of Croatia and by the
European Union through the European Regional Development Fund - the
Competitiveness and Cohesion Operational Programme (KK.01.1.1.06).

\end{document}